\documentclass[sn-mathphys-num]{sn-jnl}

\usepackage{graphicx}%
\usepackage{multirow}%
\usepackage{amsmath,amssymb,amsfonts}%
\usepackage{amsthm}%
\usepackage{mathrsfs}%
\usepackage{comment}
\usepackage[title]{appendix}%
\usepackage{xcolor}%
\usepackage{textcomp}%
\usepackage{manyfoot}%
\usepackage{booktabs}%
\usepackage{algorithm}%
\usepackage{algorithmicx}%
\usepackage{algpseudocode}%
\usepackage{listings}%

\theoremstyle{thmstyleone}%

\theoremstyle{thmstyletwo}%

\newcommand{\MSun}{\mbox{${M}_\odot$}}

\def\apgt{\ {\raise-.5ex\hbox{$\buildrel>\over\sim$}}\ }
\def\aplt{\ {\raise-.5ex\hbox{$\buildrel<\over\sim$}}\ }
\def\lteq{\ {\raise-.5ex\hbox{$\buildrel<\over-$}}\ }

\def\aap{\ {A\&A}\ }

\def\aj{\ {AJ}\ }

\def\apj{\ {ApJ}\ }

\def\mnras{\ {MNRAS}\ }

\theoremstyle{thmstylethree}%

\begin{document}

\title[]{Why Wide Jupiter-Mass Binary-Objects Cannot Form}


\author*[1]{\fnm{Simon} \sur{Portegies Zwart}}\email{spz@strw.leidenuniv.nl}
\author[1]{\fnm{Erwan} \sur{Hochart}}\email{hochart@mail.strw.leidenuniv.nl}

\affil[1]{\orgdiv{Leiden Observatory}, \orgname{Leiden University}, \orgaddress{\street{Einsteinweg}, \city{Leiden}, \postcode{2333 CC}, \country{the Netherlands}}}

\maketitle

{\bf
The discovery of $N_{\rm pp} = 40$ Jupiter-mass binary objects
(JuMBOs) alongside $N_{\rm p} = 500$ free-floating Jupiter-mass
objects (JMOs) in the Trapezium cluster's central portion raises
questions about their origin \cite{2023arXiv231001231P}.
\citet{2024NatAs...8..756W} argue that the rate at which two planets
orbiting the same star are stripped by a close encounter can explain
about half the observed JuMBOs in the Trapezium cluster.  Although,
their cross-section calculations agree with our own
\citep{2024ScPA....3....1P}, one cannot extrapolate their results into
clustered environments because it ignores the dissociation of JuMBOs
due to subsequent encounters in the clustered environment.  The
inability of forming JuMBOs via the proposed scenario either calls for
another formation mechanism, or the observed JuMBOs require thorough
confirmation.
}

\section{Introduction}\label{sec:Intro}

Particularly puzzling about the observed Jupiter-Mass Binary Objects
(JuMBOs) in the Trapezium cluster, are their wide (28\,au to 400\,au)
orbits\cite{2023arXiv231001231P}. This makes them soft pairs in the
local environment.  \citet{2024NatAs...8..756W} explored the
possibility of two Jupiter-mass planets in wide orbits around the same
host star getting knocked off their orbits after a close encounter
with a passing star. As a consequence, these stripped JMOs may form a
pair of weakly bound free-floating planets (scenario SPP for
Star-Planet-Planet, \cite{2024ScPA....3....1P}).  The results of
\cite{2024NatAs...8..756W} seem to indicate that this scenario can
explain roughly half the observed JuMBOs.  Although, their
cross-section calculations agree with our own, one cannot extrapolate
their results into clustered environments because it ignores the
softness of these systems. Integrating the SPP scenario in a clustered
environment to account for their formation as well as ionization, we
show that at best $\mathcal{O}(1)$ JuMBO is expected to be present at
any time in the Trapezium cluster.

Assuming identical Jupiter-mass objects (JMO, $m=0.001$\,\MSun\,
$\simeq 1\ {\rm M}_{\rm Jup}$) and stars ($m_\star = 1\ {\rm
  M}_\odot$), \citet{2024NatAs...8..756W} conducted $4$-body
scattering experiments to determine the JuMBO formation rate. Although
their cross-sections are consistent with other calculations
\citep{2024ScPA....3....1P,2024ApJ...970...97Y}, the extrapolation of
their results to a clustered environment lacks sustenance.

\section{Accounting for JuMBO ionisation}

For a Trapezium-like cluster environment, \citet{2024NatAs...8..756W}
find a peak $4\%$ JuMBO formation rate per star when two 
Jupiter-mass planets in planar circular orbits at 400\,au and 500\,au
(their figure 6). For any other configuration, the JuMBO production
rate drops rapidly.  A 4\% formation rate can produce 40 JuMBOs if the
cluster contains 1000 stars, and therewith explain the observed
population if they remain bound.  However, this idealised scenario,
even if all other stars would be barren, such a population would
overproduce the number of free-floating JMOs.

Assuming a broken power-law initial mass function (IMF)
\cite{2001MNRAS.322..231K}, the kinetic energy of a typical ($\langle
m_\star \rangle \sim 0.35\,M_\odot$) star in the Trapezium cluster
(with a velocity dispersion of $v_{\rm disp} = 2$\,km/s, \cite{1998ApJ...492..540H})
$>10^4$ times larger than the binding energy of the tightest observed
JuMBO.  Even free-floating JMOs in the Trapezium cluster carry $\apgt
100$\,times more kinetic energy than needed to dissociate or ionise
the tightest observed JuMBOs, making any dynamical origin improbable
and their long-term survivability questionable.

Quantifying this, the ionisation cross-section of a JuMBO with
component masses $1\ M_{\rm Jup}$ and semi-major axis $a \apgt 35$ au
is $\sigma_{\rm ion} \apgt 2 \cdot 10^{7}$ au$^{2}$ (see eq. 5.1 of
\cite{1983ApJ...268..319H}, and \cite{2024NatAs...8..756W} derive an
ionisation cross section of $5.5\times 10^{5}$\,au$^2$, but fail to
explain on how this is obtained.). The resulting ionisation timescale
in the Trapezium cluster (stellar number-density $n_\star \approx
5\times 10^{4}$ pc$^{-3}$ and $v_{\rm disp} \simeq 2$\,km/s,
\cite{1988AJ.....95.1755J}) is $\tau_{\rm ion}=1/(n_\star\sigma_{\rm
  ion} v_{\rm disp}) \approx 20$\,kyr. With a cluster age of $\sim 1$
Myr \cite{1998ApJ...492..540H} the formation timescale is $\tau_{\rm
  form} \gtrsim{(1\ {\rm Myr})}/{40} = 25$\,kyr, roughly comparable to
the ionisation timescale.  We would then expect $\sim 1$ JuMBOs to be
present in a Trapezium-like cluster at any instant, and a
substantially lower JuMBO to free-floating JMO ratio than observed.

To emphasise the improbability of JuMBO-formation through the SPP
model, we note that only $\sim 2\%$ of young stellar objects in the
Trapezium cluster host disks with sizes $r_{d}\geq 500$\,au
\cite{2005A&A...441..195V}. Given that JuMBO ionisation occurs at a
similar timescale as JuMBO formation, $\tau_{\rm ion}/\tau_{\rm form}
\sim 1$, and adopting the peak $f_{\rm peak} = 4$\,\% production rate
\cite{2024NatAs...8..756W}, then for $42$ JuMBOs to be present
simultaneously, one requires a cluster with
$N_\star\sim 42/(0.04\times0.02) \apgt 50000$ stars; about
a factor 20 larger than the observed $N_\star \approx 2500$ for the
Trapezium cluster.

\section{Extrapolating to the Trapezium Cluster}

\citet{2024NatAs...8..756W} find through scattering experiments of
isolated encounters that $f_{\rm peak}\equiv N_{\rm pp} / N_{\star}
\sim 4\%$ occurs when the outer-most JMO has orbital velocity $10$\,\%
of the encountering stars' velocity ($v_{\rm out} = 0.1v_{\rm enc}$)
and if both JMO's are on circular, co-planar orbits with an
inner-to-outer semi-major axis ratio of $a_{\rm in}/a_{\rm out} >
0.8$, and satisfies $a_{\rm out} > a_{\rm in}$.

For the Trapezium cluster, these conditions translate to $a_{\rm out}
\simeq 2.2\times10^{4}$ au, and $a_{\rm in} \sim 1.8\times10^{4}$
au. These adopted orbits are much wider than observed circumstellar
disk sizes \cite{2017ApJ...851..85B} and known planetary orbits
\cite{2024A&A...682A..43G}. In addition, two planets with such orbits
are typically separated by $<2.3$ mutual Hill radii, making the system
dynamically unstable.  Alternatively, we can assume the planetary
system to be stable (separated by 5 mutual Hill radii, or somewhat
less when in mean-motion resonance) and the two planets' orbits
separated by 100\,au\, (to reproduce the observed JuMBOs observed
projected distances). The resulting velocity dispersion when fixing
$v_{\rm out} = 0.1 v_{\rm enc}$, to correspond with the highest JuMBO
formation rate, leads to a density of $\sim 10^9$\,stars/pc$^3$, much
larger than what is observed (For $v_{\rm out}(a=500{\rm au}, M_\star
= 1.0\,\MSun) = 1.33km/s$ then leads to cluster velocity dispersion of
$v_{\rm disp} = 13.3$\,km/s, which for a 1000\,\MSun\, Plummer sphere
leads to a Plummer radius of about 0.004,pc.).  In such an
environment, the formation rate of JuMBOs is virtually zero, and their
destruction rate several orders of magnitude higher (survival
timescale is less than an orbital period).

Shifting our focus to more realistic initial conditions (between the
red-dotted and blue-dashed lines of \cite{2024NatAs...8..756W}'s
fig.\,3c), they find that one JuMBO forms for every $10^{4}$
free-floating JMOs. Although consistent with the SPP model in
\citet{2024ScPA....3....1P}, this does not agree with observations as
it can then only explain $\sim 0.13$\,\% of the observed population.
Here we assume that the identified population of 40 JuMBOs is
confirmed.  Considering complications of the spectroscopic
identification of extremely red and low-luminosity Jupiter-mass
objects in the Trapezium cluster we expect this number to drop
substantially, in which case the discrepancy with the SPP model
becomes even worse.

Ignoring the system's intrinsic instability and allowing the JMOs to
orbit nearer one another, the JuMBO-to-free-floating JMO rate
increases, yet still falls short of the observed rate: fig\,3c of
\cite{2024NatAs...8..756W} shows one JuMBO forming for every $200$
free-floating JMOs: similar to the SPM rate (for Star-Planet-Moon) of
\citet{2024ScPA....3....1P} (with which they unfortunately compare
their SPP rates while claiming consistency).  The SPM model, however,
will produce mugh tigher $\aplt 1$\,au orbits, and rather unequal
masses.

\section{Numerical Support}\label{sec:Numerical}

\begin{figure}[H]
\centering
\includegraphics[width=.69\textwidth]{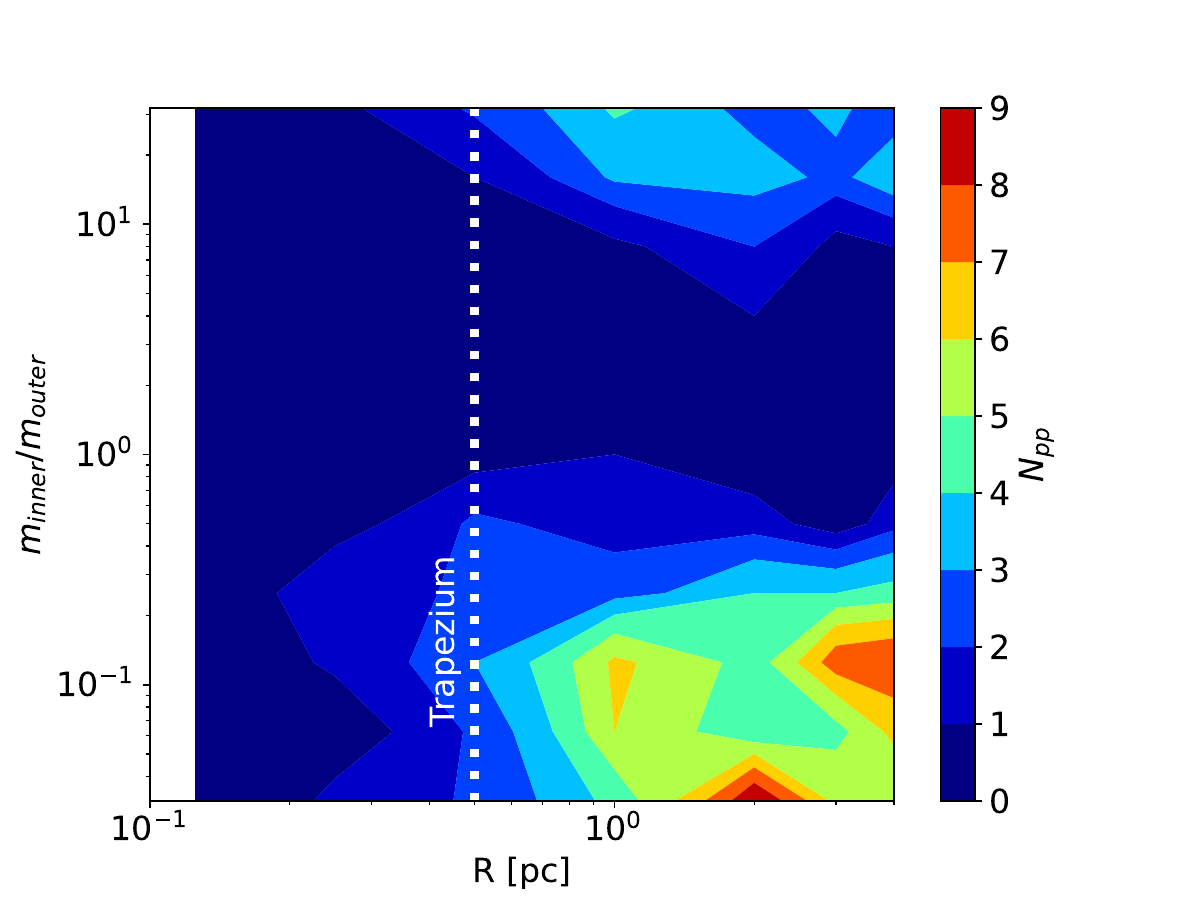}\hfill
\caption{JuMBO and free-floating JMO population at an age of
  1\,Myr. The number of JuMBOs as a function of the cluster virial
  radius ($R$) and the inner-to-outer planet mass ratio. All clusters
  are initialised with $2500$ stars from a broken power-law
  mass-function between $0.08\ M_\odot$ and $30\ M_\odot$
  \cite{2001MNRAS.322..231K} distributed in a virialised Plummer
  sphere \cite{1911MNRAS..71..460P}. We randomly select 300 stars and
  supply them with two planets, the inner planet with $m_{\rm in} =
  0.001\,M_\odot$, and the outer planet $m_{\rm out}$ (see vertical
  axis). The orbital separations are selected to ensure that $a_{\rm
    out} - a_{\rm in} = 100$\,au with the additional requirement that
  five mutual Hill-radii separate the two planets. }
\label{fig:Npp}
\end{figure}

\begin{figure}[H]
\centering
\includegraphics[width=.6\textwidth]{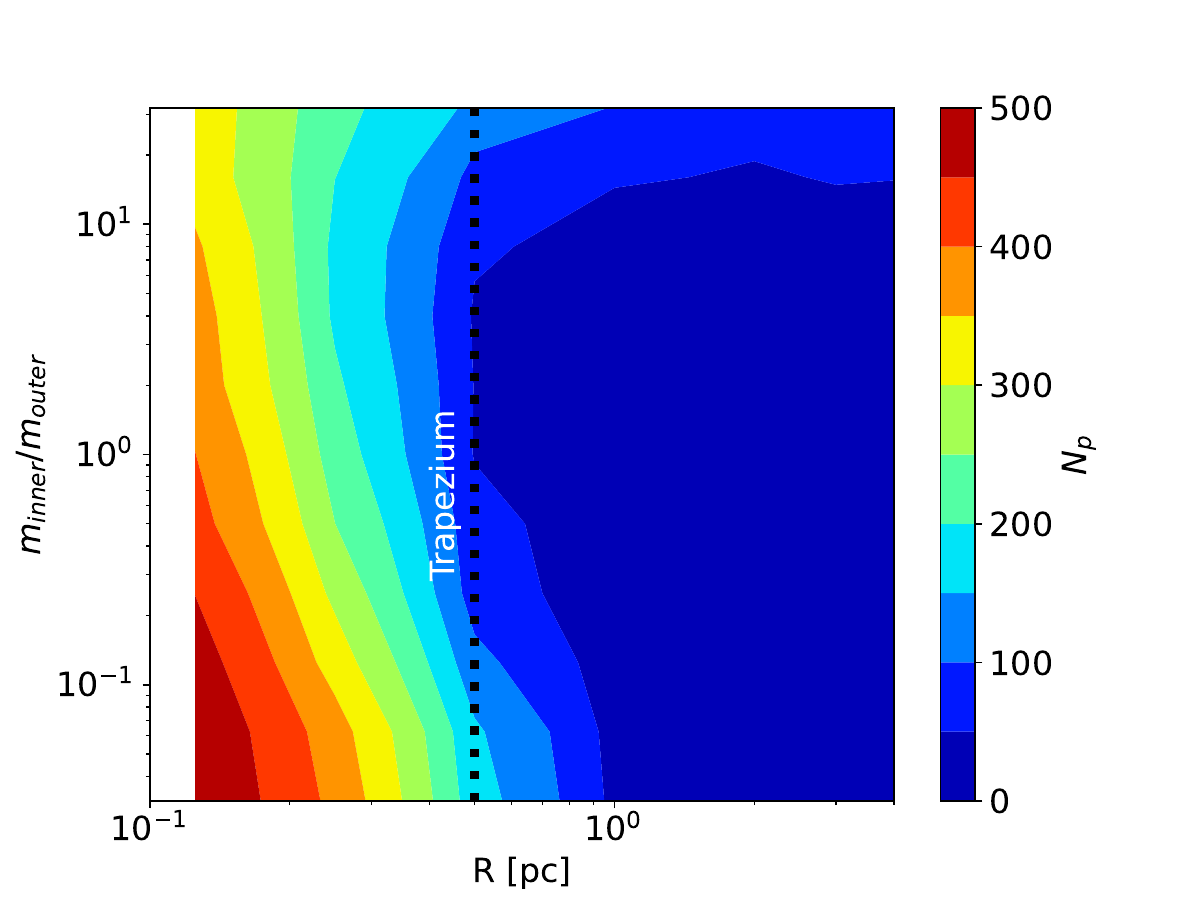}\hfill
\caption{JuMBO and free-floating JMO population at an age of
  1\,Myr. The number of free-floating Jupiter-mass objects as a
  function of the cluster virial radius ($R$) and the inner-to-outer
  planet mass ratio. Simulation parameters are identical to those
  presented in Fig.\,\ref{fig:Npp}. }
\label{fig:Np}
\end{figure}

To further support the arguments against the SPP model, we conduct new
simulations using a unsoftened 4-th order Hermite direct $N$-body code
coupled with stellar evolution through the AMUSE framework
\cite{2013CoPhC.184..456P}.  We explore both a virialised Plummer and
fractal (with fractal dimension $1.6$) distribution and evolve the
system until $1$ Myr. Fig\,\ref{fig:Npp} (for JuMBOs) and
Fig.\,\ref{fig:Np} (for JMOs) show the Plummer models' results, as the
more realistic fractal models do not produce any JuMBOs.

For Trapezium cluster's conditions \citet{2024NatAs...8..756W} find
typically $\aplt 1$ JuMBO for some $\apgt 100$ JMOs, or $\aplt 1$\%
JuMBOs among the JMOs, consistent with our finding Fig.\,\ref{fig:Npp}
and Fig.\,\ref{fig:Np}.

Decreasing $m_{\rm inner}/m_{\rm outer}$ also increases the JuMBO
formation rate. Indeed, given the right conditions where the outer JMO
is ejected, the gravitational force exerted by the outer JMO onto the
inner JMO can tip the outcome in favour of also ejecting the inner
JMO. This behaviour is more pronounced for massive outer JMOs because
of their greater gravitational influence, and leads to the increase in
JuMBOs for sparser clusters but with massive outer planets $m_m \apgt
10$\,M$_{\rm Jup}$.  Figure\,\ref{fig:Np} shows the number of rogue
JMOs liberated through dynamical encounters. The increase in rogue
planets for denser clusters is consistent with the results of
\cite{2024ScPA....3....1P} (see their table 1).

\section{Verdict}\label{sec:Verdict}

The observed number of JMO's in the Trapezium cluster could be
explained by disrupted planetary systems if half the stars had a wide
($a_{\rm out} \apgt 330$\,au) planet, but the tail of the stellar mass
function provides a more plausible explanation for the isolated sub
brown-dwarf-mass objects \citep{2006A&A...458..817W}.

We expect that at most $1$\,JuMBO could originate from the SPP model,
but then its mass ratio is probably rather small ($\aplt 0.1$).
However, the SPP model requires very wide ($a_{\rm out} \apgt
330$\,au) planetary systems to be present in the cluster. The absence
of these in the observed planetary systems does not necessary excludes
those, as such wide orbits are hard to identify in the
observations. On the other hand, only a few circum-stellar disks in
the Trapezium cluster appear to extend beyond 300\,au
\citep{2005A&A...441..195V}, and also in the Taurus association disks
appear smaller \citep{2016ApJ...827..142B}. Because of the
inefficiency of the SPP model, the lack of large disks, and the
absence of wide planetary orbits make us render the SPP model
ineffective for producing JuMBOs.

We consider the SPM scenario more promising by about an order of
magnitude, but their orbits are expected to be much tighter ($\aplt
1$\,au); which makes them still short lived (i.e., soft) at a velocity
dispersion of $\apgt 0.7$\,km/s. Overall, explaining the number of
JuMBOs and their orbital separations, either with SPP, SPM, or even
primordially, remains problematic. We can imagine that one or two
coincidence alignments appear in the data, but those systems would be
transient.

If the existence of JuMBOs is confirmed, we expect them to be either
primordial or produced by ejecting a planet-moon pair from a parent
star. In either case, they must be rare ($\aplt 10^{-3}$ per stars),
with tight ($\aplt 1$\,au) orbits, and unequal in mass.

\section{Competing interests}
The authors declare no competing interests.

\section{Data availability}
The code for this manuscript is available at\\ 
StarLab: \url{https://github.com/amusecode/Starlab}\\
The Astrophysics Multipurpose Software Environment: \url{http://amusecode.org}\\
The specific script for reproducing the runs this manuscript: \url{https://gitlab.strw.leidenuniv.nl/spz/jumboformation}.

\section{Author contributions}

\begin{itemize}
\item[] Simon Portegies Zwart: initiated the topic, wrote the run
  scripts, perform the simulations, analyzed the data, discussed the science, wrote the first
  version of the manuscript, and dealt with the refereeing and
  editorial contacts.
\item[] Erwan Hochart: initiated the topic, checked run scripts,
  performed independent validation runs, discussed the science, wrote
  the second and the version of the manuscript that eventually led to
  the submitted version.
\end{itemize}

\section{Acknowledgments}
SPZ thanks Merei for her patience in allowing the laptop to continue
running at home after office hours. We thank Mark McCaughrean, Anna
Lisa Varri, Anthony Brown, Matthew Kenworthy and Rosalba Perna for
discussions.

\section{Energy consumption}

Calculations are performed on a 13th Gen Intel Core i7-1370P (20-core
x86 64-bit Little Endian) processor, which consumes 64\, Watt. We
performed a total of 77 simulations covering cluster radius and mass
ratio. Each calculation was performed 5 times for Plummer and fractal
initial distribution functions, totaling 770 calculations of about 1
hour each. The 50\,kWh used in these calculations was produced from
solar power.

\section{Software used}

This work was made possible because of the following public software
packages, for which we are grateful to the authors:
AMUSE \cite{2018zndo...1443252P} (see \url{http://amusecode.org});
Fractal-model generator \cite{2004A&A...413..929G};
Numpy \cite{Numpy:5725236};
ph4 \cite{2022A&A...659A..86P};
pyplot \cite{2007CSE.....9...90H};
python \cite{10.5555/1593511};
SeBa \cite{1996A&A...309..179P};
Scipy \cite{SciPy};
Starlab \cite{2004MNRAS.351..473P} (see \url{https://github.com/amusecode/Starlab}).

\section{Source Data}

The scripts for generating initial conditions, performing the
calculations and analyzing the data are available at Git:
\url{https://gitlab.strw.leidenuniv.nl/spz/jumboformation}.

These scripts are based on the Astrophyiscs Multipurpose Software
Environment \cite{2018zndo...1443252P}, which is an open source
package available at \url{http://amusecode.org}.

Further data on JuMBOs is available at zenodo
\url{10.5281/zenodo.10149241}, which is produced with the source code
on github \url{https://github.com/spzwart/JuMBOs}.


\begin{thebibliography}{24}
\ifx \bisbn   \undefined \def \bisbn  #1{ISBN #1}\fi
\ifx \binits  \undefined \def \binits#1{#1}\fi
\ifx \bauthor  \undefined \def \bauthor#1{#1}\fi
\ifx \batitle  \undefined \def \batitle#1{#1}\fi
\ifx \bjtitle  \undefined \def \bjtitle#1{#1}\fi
\ifx \bvolume  \undefined \def \bvolume#1{\textbf{#1}}\fi
\ifx \byear  \undefined \def \byear#1{#1}\fi
\ifx \bissue  \undefined \def \bissue#1{#1}\fi
\ifx \bfpage  \undefined \def \bfpage#1{#1}\fi
\ifx \blpage  \undefined \def \blpage #1{#1}\fi
\ifx \burl  \undefined \def \burl#1{\textsf{#1}}\fi
\ifx \doiurl  \undefined \def \doiurl#1{\url{https://doi.org/#1}}\fi
\ifx \betal  \undefined \def \betal{\textit{et al.}}\fi
\ifx \binstitute  \undefined \def \binstitute#1{#1}\fi
\ifx \binstitutionaled  \undefined \def \binstitutionaled#1{#1}\fi
\ifx \bctitle  \undefined \def \bctitle#1{#1}\fi
\ifx \beditor  \undefined \def \beditor#1{#1}\fi
\ifx \bpublisher  \undefined \def \bpublisher#1{#1}\fi
\ifx \bbtitle  \undefined \def \bbtitle#1{#1}\fi
\ifx \bedition  \undefined \def \bedition#1{#1}\fi
\ifx \bseriesno  \undefined \def \bseriesno#1{#1}\fi
\ifx \blocation  \undefined \def \blocation#1{#1}\fi
\ifx \bsertitle  \undefined \def \bsertitle#1{#1}\fi
\ifx \bsnm \undefined \def \bsnm#1{#1}\fi
\ifx \bsuffix \undefined \def \bsuffix#1{#1}\fi
\ifx \bparticle \undefined \def \bparticle#1{#1}\fi
\ifx \barticle \undefined \def \barticle#1{#1}\fi
\bibcommenthead
\ifx \bconfdate \undefined \def \bconfdate #1{#1}\fi
\ifx \botherref \undefined \def \botherref #1{#1}\fi
\ifx \url \undefined \def \url#1{\textsf{#1}}\fi
\ifx \bchapter \undefined \def \bchapter#1{#1}\fi
\ifx \bbook \undefined \def \bbook#1{#1}\fi
\ifx \bcomment \undefined \def \bcomment#1{#1}\fi
\ifx \oauthor \undefined \def \oauthor#1{#1}\fi
\ifx \citeauthoryear \undefined \def \citeauthoryear#1{#1}\fi
\ifx \endbibitem  \undefined \def \endbibitem {}\fi
\ifx \bconflocation  \undefined \def \bconflocation#1{#1}\fi
\ifx \arxivurl  \undefined \def \arxivurl#1{\textsf{#1}}\fi
\csname PreBibitemsHook\endcsname

\bibitem[\protect\citeauthoryear{{Pearson} and
  {McCaughrean}}{2023}]{2023arXiv231001231P}
\begin{botherref}
\oauthor{\bsnm{{Pearson}}, \binits{S.G.}},
\oauthor{\bsnm{{McCaughrean}}, \binits{M.J.}}:
{Jupiter Mass Binary Objects in the Trapezium Cluster}.
arXiv e-prints,
2310--01231
(2023)
\doiurl{10.48550/arXiv.2310.01231}
{\href{https://arxiv.org/abs/2310.01231}{{arXiv:2310.01231}}}
{[astro-ph.EP]}
\end{botherref}
\endbibitem

\bibitem[\protect\citeauthoryear{{Wang} et~al.}{2024}]{2024NatAs...8..756W}
\begin{barticle}
\bauthor{\bsnm{{Wang}}, \binits{Y.}},
\bauthor{\bsnm{{Perna}}, \binits{R.}},
\bauthor{\bsnm{{Zhu}}, \binits{Z.}}:
\batitle{{Free-floating binary planets from ejections during close stellar
  encounters}}.
\bjtitle{Nature Astronomy}
\bvolume{8},
\bfpage{756}--\blpage{764}
(\byear{2024})
\doiurl{10.1038/s41550-024-02239-2}
{\href{https://arxiv.org/abs/2310.06016}{{arXiv:2310.06016}}}
{[astro-ph.EP]}
\end{barticle}
\endbibitem

\bibitem[\protect\citeauthoryear{{Portegies Zwart} and
  {Hochart}}{2024}]{2024ScPA....3....1P}
\begin{barticle}
\bauthor{\bsnm{{Portegies Zwart}}, \binits{S.}},
\bauthor{\bsnm{{Hochart}}, \binits{E.}}:
\batitle{{The origin and evolution of wide Jupiter mass binary objects in young
  stellar clusters}}.
\bjtitle{SciPost Astronomy}
\bvolume{3}(\bissue{1}),
\bfpage{001}
(\byear{2024})
\doiurl{10.21468/SciPostAstro.3.1.001}
\end{barticle}
\endbibitem

\bibitem[\protect\citeauthoryear{{Yu} and {Lai}}{2024}]{2024ApJ...970...97Y}
\begin{barticle}
\bauthor{\bsnm{{Yu}}, \binits{F.}},
\bauthor{\bsnm{{Lai}}, \binits{D.}}:
\batitle{{Free-floating Planets, Survivor Planets, Captured Planets, and Binary
  Planets from Stellar Flybys}}.
\bjtitle{\apj}
\bvolume{970}(\bissue{1}),
\bfpage{97}
(\byear{2024})
\doiurl{10.3847/1538-4357/ad4f81}
{\href{https://arxiv.org/abs/2403.07224}{{arXiv:2403.07224}}}
{[astro-ph.EP]}
\end{barticle}
\endbibitem

\bibitem[\protect\citeauthoryear{{Kroupa}}{2001}]{2001MNRAS.322..231K}
\begin{barticle}
\bauthor{\bsnm{{Kroupa}}, \binits{P.}}:
\batitle{{On the variation of the initial mass function}}.
\bjtitle{\mnras}
\bvolume{322},
\bfpage{231}--\blpage{246}
(\byear{2001})
\end{barticle}
\endbibitem

\bibitem[\protect\citeauthoryear{{Hillenbrand} and
  {Hartmann}}{1998}]{1998ApJ...492..540H}
\begin{barticle}
\bauthor{\bsnm{{Hillenbrand}}, \binits{L.A.}},
\bauthor{\bsnm{{Hartmann}}, \binits{L.W.}}:
\batitle{{A Preliminary Study of the Orion Nebula Cluster Structure and
  Dynamics}}.
\bjtitle{\apj}
\bvolume{492},
\bfpage{540}--\blpage{553}
(\byear{1998})
\doiurl{10.1086/305076}
\end{barticle}
\endbibitem

\bibitem[\protect\citeauthoryear{{Hut} and
  {Bahcall}}{1983}]{1983ApJ...268..319H}
\begin{barticle}
\bauthor{\bsnm{{Hut}}, \binits{P.}},
\bauthor{\bsnm{{Bahcall}}, \binits{J.N.}}:
\batitle{{Binary-single star scattering. I - Numerical experiments for equal
  masses}}.
\bjtitle{\apj}
\bvolume{268},
\bfpage{319}--\blpage{341}
(\byear{1983})
\doiurl{10.1086/160956}
\end{barticle}
\endbibitem

\bibitem[\protect\citeauthoryear{{Jones} and
  {Walker}}{1988}]{1988AJ.....95.1755J}
\begin{barticle}
\bauthor{\bsnm{{Jones}}, \binits{B.F.}},
\bauthor{\bsnm{{Walker}}, \binits{M.F.}}:
\batitle{{Proper Motions and Variabilities of Stars Near the Orion Nebula}}.
\bjtitle{\aj}
\bvolume{95},
\bfpage{1755}
(\byear{1988})
\doiurl{10.1086/114773}
\end{barticle}
\endbibitem

\bibitem[\protect\citeauthoryear{{Vicente} and
  {Alves}}{2005}]{2005A&A...441..195V}
\begin{barticle}
\bauthor{\bsnm{{Vicente}}, \binits{S.M.}},
\bauthor{\bsnm{{Alves}}, \binits{J.}}:
\batitle{{Size distribution of circumstellar disks in the Trapezium cluster}}.
\bjtitle{\aap}
\bvolume{441},
\bfpage{195}--\blpage{205}
(\byear{2005})
\doiurl{10.1051/0004-6361:20053540}
{\href{https://arxiv.org/abs/astro-ph/0506585}{{astro-ph/0506585}}}
\end{barticle}
\endbibitem

\bibitem[\protect\citeauthoryear{{Barenfeld} et~al.}{2017}]{2017ApJ...851..85B}
\begin{barticle}
\bauthor{\bsnm{{Barenfeld}}, \binits{S.A.}},
\bauthor{\bsnm{{Carpenter}}, \binits{J.M.}},
\bauthor{\bsnm{{Sargent}}, \binits{A.I.}},
\bauthor{\bsnm{{Isella}}, \binits{A.}},
\bauthor{\bsnm{{Ricci}}, \binits{L.}}:
\batitle{{Measurement of Circumstellar Disk Sizes in the Upper Scorpius OB
  Association with ALMA}}.
\bjtitle{\apj}
\bvolume{851}(\bissue{2}),
\bfpage{85}
(\byear{2017})
\doiurl{10.3847/1538-4357/aa989d}
{\href{https://arxiv.org/abs/1711.04045}{{arXiv:1711.04045}}}
{[astro-ph.SR]}
\end{barticle}
\endbibitem

\bibitem[\protect\citeauthoryear{{Gurrutxaga}
  et~al.}{2024}]{2024A&A...682A..43G}
\begin{barticle}
\bauthor{\bsnm{{Gurrutxaga}}, \binits{N.}},
\bauthor{\bsnm{{Johansen}}, \binits{A.}},
\bauthor{\bsnm{{Lambrechts}}, \binits{M.}},
\bauthor{\bsnm{{Appelgren}}, \binits{J.}}:
\batitle{{Formation of wide-orbit giant planets in protoplanetary disks with a
  decreasing pebble flux}}.
\bjtitle{\aap}
\bvolume{682},
\bfpage{43}
(\byear{2024})
\doiurl{10.1051/0004-6361/202348020}
{\href{https://arxiv.org/abs/2311.04365}{{arXiv:2311.04365}}}
{[astro-ph.EP]}
\end{barticle}
\endbibitem

\bibitem[\protect\citeauthoryear{{Portegies Zwart}
  et~al.}{2013}]{2013CoPhC.184..456P}
\begin{barticle}
\bauthor{\bsnm{{Portegies Zwart}}, \binits{S.F.}},
\bauthor{\bsnm{{McMillan}}, \binits{S.L.W.}},
\bauthor{\bsnm{{van Elteren}}, \binits{A.}},
\bauthor{\bsnm{{Pelupessy}}, \binits{F.I.}},
\bauthor{\bsnm{{de Vries}}, \binits{N.}}:
\batitle{{Multi-physics simulations using a hierarchical interchangeable
  software interface}}.
\bjtitle{Computer Physics Communications}
\bvolume{184},
\bfpage{456}--\blpage{468}
(\byear{2013})
\doiurl{10.1016/j.cpc.2012.09.024}
\end{barticle}
\endbibitem

\bibitem[\protect\citeauthoryear{{Plummer}}{1911}]{1911MNRAS..71..460P}
\begin{barticle}
\bauthor{\bsnm{{Plummer}}, \binits{H.C.}}:
\batitle{{On the problem of distribution in globular star clusters}}.
\bjtitle{\mnras}
\bvolume{71},
\bfpage{460}--\blpage{470}
(\byear{1911})
\end{barticle}
\endbibitem

\bibitem[\protect\citeauthoryear{{Whitworth} and
  {Stamatellos}}{2006}]{2006A&A...458..817W}
\begin{barticle}
\bauthor{\bsnm{{Whitworth}}, \binits{A.P.}},
\bauthor{\bsnm{{Stamatellos}}, \binits{D.}}:
\batitle{{The minimum mass for star formation, and the origin of binary brown
  dwarfs}}.
\bjtitle{\aap}
\bvolume{458}(\bissue{3}),
\bfpage{817}--\blpage{829}
(\byear{2006})
\doiurl{10.1051/0004-6361:20065806}
{\href{https://arxiv.org/abs/astro-ph/0610039}{{arXiv:astro-ph/0610039}}}
{[astro-ph]}
\end{barticle}
\endbibitem

\bibitem[\protect\citeauthoryear{{Barenfeld}
  et~al.}{2016}]{2016ApJ...827..142B}
\begin{barticle}
\bauthor{\bsnm{{Barenfeld}}, \binits{S.A.}},
\bauthor{\bsnm{{Carpenter}}, \binits{J.M.}},
\bauthor{\bsnm{{Ricci}}, \binits{L.}},
\bauthor{\bsnm{{Isella}}, \binits{A.}}:
\batitle{{ALMA Observations of Circumstellar Disks in the Upper Scorpius OB
  Association}}.
\bjtitle{\apj}
\bvolume{827}(\bissue{2}),
\bfpage{142}
(\byear{2016})
\doiurl{10.3847/0004-637X/827/2/142}
{\href{https://arxiv.org/abs/1605.05772}{{arXiv:1605.05772}}}
{[astro-ph.EP]}
\end{barticle}
\endbibitem

\bibitem[\protect\citeauthoryear{{Portegies Zwart}
  et~al.}{2018}]{2018zndo...1443252P}
\begin{botherref}
\oauthor{\bsnm{{Portegies Zwart}}, \binits{S.}},
\oauthor{\bsnm{{van Elteren}}, \binits{A.}},
\oauthor{\bsnm{{Pelupessy}}, \binits{I.}},
\oauthor{\bsnm{{McMillan}}, \binits{S.}},
\oauthor{\bsnm{{Rieder}}, \binits{S.}},
\oauthor{\bsnm{{de Vries}}, \binits{N.}},
\oauthor{\bsnm{{Marosvolgyi}}, \binits{M.}},
\oauthor{\bsnm{{Whitehead}}, \binits{A.}},
\oauthor{\bsnm{{Wall}}, \binits{J.}},
\oauthor{\bsnm{{Drost}}, \binits{N.}},
\oauthor{\bsnm{{J{\'\i}lkov{\'a}}}, \binits{L.}},
\oauthor{\bsnm{{Martinez Barbosa}}, \binits{C.}},
\oauthor{\bsnm{{van der Helm}}, \binits{E.}},
\oauthor{\bsnm{{Beedorf}}, \binits{J.}},
\oauthor{\bsnm{{Bos}}, \binits{P.}},
\oauthor{\bsnm{{Boekholt}}, \binits{T.}},
\oauthor{\bsnm{{van Werkhoven}}, \binits{B.}},
\oauthor{\bsnm{{Wijnen}}, \binits{T.}},
\oauthor{\bsnm{{Hamers}}, \binits{A.}},
\oauthor{\bsnm{{Caputo}}, \binits{D.}},
\oauthor{\bsnm{{Ferrari}}, \binits{G.}},
\oauthor{\bsnm{{Toonen}}, \binits{S.}},
\oauthor{\bsnm{{Gaburov}}, \binits{E.}},
\oauthor{\bsnm{{Paardekooper}}, \binits{J.-P.}},
\oauthor{\bsnm{{Janes}}, \binits{J.}},
\oauthor{\bsnm{{Punzo}}, \binits{D.}},
\oauthor{\bsnm{{Kruip}}, \binits{C.}},
\oauthor{\bsnm{{Altay}}, \binits{G.}}:
{Amuse: The Astrophysical Multipurpose Software Environment}.
Zenodo
(2018).
\doiurl{10.5281/zenodo.1443252}
\end{botherref}
\endbibitem

\bibitem[\protect\citeauthoryear{{Goodwin} and
  {Whitworth}}{2004}]{2004A&A...413..929G}
\begin{barticle}
\bauthor{\bsnm{{Goodwin}}, \binits{S.P.}},
\bauthor{\bsnm{{Whitworth}}, \binits{A.P.}}:
\batitle{{The dynamical evolution of fractal star clusters: The survival of
  substructure}}.
\bjtitle{\aap}
\bvolume{413},
\bfpage{929}--\blpage{937}
(\byear{2004})
\doiurl{10.1051/0004-6361:20031529}
{\href{https://arxiv.org/abs/astro-ph/0310333}{{astro-ph/0310333}}}
\end{barticle}
\endbibitem

\bibitem[\protect\citeauthoryear{van~der Walt et~al.}{2011}]{Numpy:5725236}
\begin{barticle}
\bauthor{\bsnm{Walt}, \binits{S.}},
\bauthor{\bsnm{Colbert}, \binits{S.C.}},
\bauthor{\bsnm{Varoquaux}, \binits{G.}}:
\batitle{The numpy array: A structure for efficient numerical computation}.
\bjtitle{Computing in Science Engineering}
\bvolume{13}(\bissue{2}),
\bfpage{22}--\blpage{30}
(\byear{2011})
\doiurl{10.1109/MCSE.2011.37}
\end{barticle}
\endbibitem

\bibitem[\protect\citeauthoryear{{Portegies Zwart}
  et~al.}{2022}]{2022A&A...659A..86P}
\begin{barticle}
\bauthor{\bsnm{{Portegies Zwart}}, \binits{S.F.}},
\bauthor{\bsnm{{Boekholt}}, \binits{T.C.N.}},
\bauthor{\bsnm{{Por}}, \binits{E.H.}},
\bauthor{\bsnm{{Hamers}}, \binits{A.S.}},
\bauthor{\bsnm{{McMillan}}, \binits{S.L.W.}}:
\batitle{{Chaos in self-gravitating many-body systems. Lyapunov time dependence
  of N and the influence of general relativity}}.
\bjtitle{\aap}
\bvolume{659},
\bfpage{86}
(\byear{2022})
\doiurl{10.1051/0004-6361/202141789}
{\href{https://arxiv.org/abs/2109.11012}{{arXiv:2109.11012}}}
{[nlin.CD]}
\end{barticle}
\endbibitem

\bibitem[\protect\citeauthoryear{{Hunter}}{2007}]{2007CSE.....9...90H}
\begin{barticle}
\bauthor{\bsnm{{Hunter}}, \binits{J.D.}}:
\batitle{{Matplotlib: A 2D Graphics Environment}}.
\bjtitle{Computing in Science and Engineering}
\bvolume{9},
\bfpage{90}--\blpage{95}
(\byear{2007})
\doiurl{10.1109/MCSE.2007.55}
\end{barticle}
\endbibitem

\bibitem[\protect\citeauthoryear{Van~Rossum and Drake}{2009}]{10.5555/1593511}
\begin{bbook}
\bauthor{\bsnm{Van~Rossum}, \binits{G.}},
\bauthor{\bsnm{Drake}, \binits{F.L.}}:
\bbtitle{Python 3 Reference Manual}.
\bpublisher{CreateSpace},
\blocation{Scotts Valley, CA}
(\byear{2009})
\end{bbook}
\endbibitem

\bibitem[\protect\citeauthoryear{{Portegies Zwart} and
  {Verbunt}}{1996}]{1996A&A...309..179P}
\begin{barticle}
\bauthor{\bsnm{{Portegies Zwart}}, \binits{S.F.}},
\bauthor{\bsnm{{Verbunt}}, \binits{F.}}:
\batitle{{Population synthesis of high-mass binaries.}}
\bjtitle{\aap}
\bvolume{309},
\bfpage{179}--\blpage{196}
(\byear{1996})
\end{barticle}
\endbibitem

\bibitem[\protect\citeauthoryear{Jones et~al.}{2001--}]{SciPy}
\begin{botherref}
\oauthor{\bsnm{Jones}, \binits{E.}},
\oauthor{\bsnm{Oliphant}, \binits{T.}},
\oauthor{\bsnm{Peterson}, \binits{P.}}, et al.:
{SciPy}: Open source scientific tools for {Python}
(2001--).
\url{http://www.scipy.org/}
\end{botherref}
\endbibitem

\bibitem[\protect\citeauthoryear{{Portegies Zwart}
  et~al.}{2004}]{2004MNRAS.351..473P}
\begin{barticle}
\bauthor{\bsnm{{Portegies Zwart}}, \binits{S.F.}},
\bauthor{\bsnm{{Hut}}, \binits{P.}},
\bauthor{\bsnm{{McMillan}}, \binits{S.L.W.}},
\bauthor{\bsnm{{Makino}}, \binits{J.}}:
\batitle{{Star cluster ecology - V. Dissection of an open star cluster:
  spectroscopy}}.
\bjtitle{\mnras}
\bvolume{351},
\bfpage{473}--\blpage{486}
(\byear{2004})
\doiurl{10.1111/j.1365-2966.2004.07709.x}
{\href{https://arxiv.org/abs/arXiv:astro-ph/0301041}{{arXiv:astro-ph/0301041}}}
\end{barticle}
\endbibitem

\end{thebibliography}


\end{document}